\documentclass[preprint,showpacs,
amssymb]{revtex4}
\usepackage{graphicx}
\usepackage{dcolumn}
\usepackage{bm}
\begin{document}

\title{The $f_2$- and $\rho_3$-mesons in multi-channel pion-pion scattering}

\author{Yu.S.~Surovtsev\footnote{E-mail address: surovcev@theor.jinr.ru}}
\affiliation {Bogoliubov Laboratory of Theoretical Physics, JINR,
Dubna 141980, Russia}
\author{P.~Byd\v{z}ovsk\'y\footnote{E-mail address: bydz@ujf.cas.cz}}
\affiliation {Nuclear Physics Institute, ASCR, \v{R}e\v{z} near
Prague 25068, Czech Republic}
\author{R.~Kami\'nski\footnote{E-mail address: Robert.Kaminski@ifj.edu.pl}}
\affiliation {Institute of Nuclear Physics, PAN, Cracow 31342,
Poland}
\author{M.~Nagy\footnote{E-mail address: fyzinami@unix.savba.sk}}
\affiliation {Inst. Phys. Slovak Acad. Sci., Dubravska cesta 9,
 845 11 Bratislava, Slovak Republic}

\date{\today}

\begin{abstract}
In a multi-channel $S$-matrix approach, data on
$\pi\pi\to\pi\pi,K\overline{K},\eta\eta$ in the
$I^GJ^{PC}=0^+2^{++}$ sector and $\pi\pi$ scattering in the
$1^+3^{--}$ sector are analyzed to study the $f_2$- and
$\rho_3$-mesons, respectively. Spectroscopic implications and
possible classification of the $f_2$-states in terms of the SU(3)
multiplets are also discussed.

\end{abstract}
\maketitle

\section{Introduction}

We present results of the coupled-channel analysis of data on
processes $\pi\pi\to\pi\pi,K\overline{K},\eta\eta$ in the
$I^GJ^{PC}=0^+2^{++}$ sector and on the $\pi\pi$ scattering in the
$1^+3^{--}$ sector.

Our knowledge about the existence and parameters of resonances
in the $0^+2^{++}$ sector is not clear yet.
Nine from the thirteen resonances, discussed in the PDG
issue \cite{PDG10} and in the literature \cite{Anis05},
$f_2(1430)$, $f_2(1565)$, $f_2(1640)$, $f_2(1810)$, $f_2(1910)$,
$f_2(2000)$, $f_2(2020)$, $f_2(2150)$, $f_2(2220)$, must be still
confirmed in various experiments and analyses.
In the analysis of processes
$p\overline{p}\to\pi\pi,\eta\eta,\eta\eta^\prime$ \cite{Anis05}
five resonances -- $f_2(1920)$, $f_2(2000)$, $f_2(2020)$,
$f_2(2240)$ and $f_2(2300)$ -- have been obtained, where the
$f_2(2000)$ is a candidate for the glueball. In our analysis of
$\pi\pi\to\pi\pi,K\overline{K},\eta\eta$ \cite{SBKN-PRD10} we
supported this conclusion on the $f_2(2000)$.

The tensor sector is also interesting because here multi-quark
states might be observed apparently as separate states, which are
difficult to observe in the scalar sector where owing to their
large widths these states can manifest themselves only in a
distortion of the $q{\bar q}$ picture.

Investigation in the $I^GJ^{PC}=1^+3^{--}$ sector is motivated by
those results \cite{SBKN-PRD10,SBGL-ppn10} in the $0^+2^{++}$,
$0^+0^{++}$ and $1^+1^{--}$ sectors, which (if they are confirmed)
will require revisions of the mainstream quark models, e.g.
\cite{Isg-85}, and by a possibility to support (or not) these
results when studying of other mesonic sectors. These are the
earlier obtained disagreements with predictions of the indicated
model, e.g., with respect to the $f_0(600)$ and $f_0(1500)$ in the
scalar sector and to the second $q{\bar q}$ nonet in the tensor
sector \cite{SBKN-PRD10,SBGL-ppn10}.
Especially it is worth to remind that result in the vector sector:
In our multi-channel analysis \cite{SBKN-PRD10,SB-08} of the
$P$-wave $\pi\pi$ scattering data \cite{v-pipi} and in the
re-analysis of the process $e^+e^-\to\omega\pi^0$ \cite{Yam-07},
the old conclusion \cite{Bud-77} was confirmed (which, to the point,
is consistent with results of some quark models \cite{GG-82}) that
the first $\rho$-like meson is $\rho(1250)$ unlike $\rho(1450)$
cited in the PDG tables \cite{PDG10}. However, existence of both
states does not contradict to the $\pi\pi$ scattering data
\cite{SBKN-PRD10,SB-08}. It is important that for both states there
are apparently possible SU(3) partners. For the $\rho(1250)$ these
partners are: the isodoublet $K^*(1410)$ and the isoscalar
$\omega(1420)$, for which the obtained mass is in range
1350-1460~MeV \cite{PDG10}, whereas the Gell-Mann--Okubo (GM-O)
formula
$$3m_{\omega_8^\prime}^2=4m_{{K^*}^\prime}^2-m_{\rho^\prime}^2$$
gives for the mass of the eighth component of corresponding octet
the value about 1460~MeV. The $\rho(1450)$, which might be the
isovector $^3D_1$ state in the $q{\bar q}$ picture, could be put
into the octet together with the isodoublet $K^*(1680)$. Then from
the GM-O formula, the value 1750~MeV is obtained for the mass of
the eighth component of this octet. This corresponds to one of the
observations of the second $\omega$-like meson which is cited in
the PDG tables under the $\omega(1650)$ and has the mass, obtained
in various works, from 1606 to 1840~MeV.

In the mainstream quark model \cite{Isg-85}, the first $\rho$-like
meson is usually predicted by about 200~MeV higher than the
$\rho(1250)$, and also the first $K^*$-like meson is obtained at
1580~MeV, whereas the corresponding well established resonance has
the mass of about 1410~MeV.
Therefore, it is important to check if the conclusion on the
$\rho(1250)$ is supported by investigation in other mesonic sectors.
Considering the $(J,M^2)$-plot for the daughter $\rho$-trajectory,
related to the suggested $\rho(1250)$, one concludes that there
should exist the $1^+3^{--}$-state at about 1950~MeV
--``$\rho_3(1950)$''. It is worth to check this state analyzing
accessible data on the $F$-wave $\pi\pi$ scattering \cite{v-pipi}.

In the present investigation, we applied the multi-channel
$S$-matrix approach \cite{SBKN-PRD10}. To generate resonance poles
and zeros on the Riemann surface, we used the multi-channel
Breit--Wigner forms taking into account the Blatt--Weisskopf
barrier factors given by spins of resonances \cite{Blatt-52}.

\section{The $S$-matrix formalism for N coupled channels}

The N-channel $S$-matrix is determined on the $2^N$-sheeted
Riemann surface. The matrix elements $S_{ab}$~ ($a, b = 1, 2,\cdots, N$
denote channels) have the right-hand cuts along the real axis of
the complex-$s$ plane ($s$ is the invariant total energy squared),
related to the considered channels and starting in the channel
thresholds $s_i$ ($i=1,\cdots, N$), and the left-hand cuts related
to the crossed channels. The main model-independent part of
resonance contributions is given by poles and zeros on the Riemann
surface. Generally, this representation of resonances can be obtained
utilizing formulas for the analytic continuations of the matrix
elements for the coupled processes to the unphysical sheets of the
Riemann surface, as it was performed for the N-channel case in
Ref.~\cite{KMS-nc96}.

In this work, the Le Couteur--Newton relations \cite{LeCN} are used to
generate the resonance poles and zeros on the Riemann surface. These
relations express the $S$-matrix elements of all coupled processes
in terms of the Jost matrix determinant $d(k_1,\cdots,k_N)$
($k_i=\frac{1}{2}\sqrt{s-s_i}$) that is a real analytic function
with the only branch-points at $k_i=0$: \vspace*{-0.2cm}
\begin{equation}
S_{aa}=\frac{d(k_1,\cdots,k_{a-1},-k_a,k_{a+1},\cdots,k_N)}
{d(k_1,\cdots,k_N)},$$
$$S_{aa}S_{bb}-S_{ab}^2=\frac{d(k_1,\cdots,k_{a-1},-k_a,k_{a+1},
\cdots,k_{b-1},-k_b,k_{b+1}, \cdots,k_N)}{d(k_1,\cdots,k_N)}.
\end{equation}
The real analyticity implies~~~ $d(s^*)=d^*(s)~~~~~~{\rm for~~
all}~~ s$. The N-channel unitarity requires
$$|d(k_1,\cdots,-k_a,\cdots,k_N)|\leq |d(k_1,\cdots,k_N)|,~
a=1,\cdots,N,$$
$$|d(-k_1,\cdots,-k_a,\cdots,-k_N)|=
|d(k_1,\cdots,k_a,\cdots,k_N)|$$ to hold for physical values of $s$.

The $d$-function is taken in the separable form ~$d=d_B d_{res}$.
The resonance part $d_{res}$ is described using the multi-channel
Breit--Wigner forms
\begin{equation} \label{d_{res}}
d_{res}(s)=\prod_{r}
\left[M_r^2-s-i\sum_{i=1}^N\rho_{ri}^{2J+1}R_{ri}f_{ri}^2\right]\,,
\end{equation}
where $\rho_{ri}=2k_i/\sqrt{M_r^2-s_i}\,$,~ $f_{ri}^2/M_r$ indicates
to the partial width of a resonance with mass $M_r$, and
$R_{ri}(s,M_r,s_i,r_{ri})$ are the Blatt--Weisskopf barrier
factors with $s_i$ the channel threshold and
$r_{ri}$ a radius of the $i$-channel decay of the state ``$r$''.

The background part $d_B$ represents mainly an influence of channels
which are not explicitly included. Opening of these channels causes
a rise of the corresponding elastic and inelastic phase shifts
in $d_B$  \vspace*{-0.2cm}
\begin{equation} \label{d_B}
d_B=\mbox{exp}\left[-i\sum_{i=1}^{N}\left(\sqrt{\frac{s-s_i}{s}}
\right)^{2J+1}(a_i + ib_i)\right].
\end{equation}

From the formulas of analytic continuation of the matrix elements for
the coupled processes to the unphysical sheets of the Riemann surface
\cite{KMS-nc96}, one can conclude that only on the sheets with the
numbers $2^i$ ($i=1,\cdots,N$), i.e. II, IV, VIII, XVI,..., the
analytic continuations have the form $\propto1/S_{ii}^{\rm I}$ where
$S_{ii}^{\rm I}$ is the $S$-matrix element on the physical (I) sheet.
This means that only on these sheets the pole positions of resonances
are at the same points of the $s$-plane, as the resonance zeros on the
physical sheet, i.e., they are not shifted due to the coupling of channels.
Therefore, {\it the resonance parameters should be calculated from the
pole positions only on these sheets.}

In the four-channel case, considered below, the Riemann surface is
sixteen-sheeted. The sheets II, IV, VIII, and XVI correspond to
the following signs of analytic continuations of the quantities
${\mbox{Im}}\sqrt{s-s_1}$, ${\mbox{Im}}\sqrt{s-s_2}$,
${\mbox{Im}}\sqrt{s-s_3}$, and ${\mbox{Im}}\sqrt{s-s_4}$:
$-+++,+-++,++-+$, and $+++-$, respectively.

\section{Analysis of the $I^GJ^{PC}=0^+2^{++}$ sector}

In the analysis of data on the isoscalar D-waves of processes
$\pi\pi\!\to\!\pi\pi,K\overline{K},\eta\eta$, we have considered explicitly
also the channel $(2\pi)(2\pi)$. Therefore, we have applied the
four-channel Breit--Wigner form for the resonance part (\ref{d_{res}}) of
the function $d(\sqrt{s-s_1},\sqrt{s-s_2},\sqrt{s-s_3},\sqrt{s-s_4})$.
The Blatt--Weisskopf barrier factor for a particle with $J=2$ is
\begin{equation} \label{R_{ri}}
R_{ri}=\frac{9+\frac{3}{4}
(\sqrt{M_r^2-s_i}~r_{ri})^2+\frac{1}{16}(\sqrt{M_r^2-s_i}~r_{ri})^4}
{9+\frac{3}{4}(\sqrt{s-s_i}~r_{ri})^2+\frac{1}{16}(\sqrt{s-s_i}~r_{ri})^4}
\end{equation}
with radii 0.943 fm for resonances in all channels, except
for $f_2(1270)$ and $f_2(1960)$ for which the radii are
(as the results of the analysis): for $f_2(1270)$,~ 1.498, 0.708 and
0.606 fm in channels $\pi\pi$, $K\overline{K}$ and $\eta\eta$,
respectively, and for $f_2(1960)$, ~0.296 fm in channel $K\overline{K}$.

The background part (\ref{d_B}) has the form
\begin{equation}
d_B=\mbox{exp}\left[-i\sum_{n=1}^{3}\left(\sqrt{\frac{s-s_n}{s}}
\right)^5(a_n+ ib_n)\right]\,,
\end{equation}
where
$$
a_1=\alpha_{11}+\frac{s-4m_K^2}{s}~\alpha_{12}~\theta(s-4m_K^2)+
\frac{s-s_v}{s}~\alpha_{10}~\theta(s-s_v),
$$
$$
b_n=\beta_n+\frac{s-s_v}{s}~\gamma_n~\theta(s-s_v).
$$
$s_v\approx2.274$ GeV$^2$ is a combined threshold of the channels
$\eta\eta^{\prime},~ \rho\rho,$ and $\omega\omega$.

The data for the $\pi\pi$ scattering are taken from an energy-independent
analysis by B.~Hyams et al. \cite{v-pipi}. The data for
$\pi\pi\to K\overline{K},\eta\eta$ are taken from works \cite{Lind92}.

A satisfactory description (with the total
$\chi^2/\mbox{NDF}=161.147/(168-65)\approx 1.56$) is obtained both
with ten resonances -- $f_2(1270)$, $f_2(1430)$, $f_2^{\prime}(1525)$,
$f_2(1580)$, $f_2(1730)$, $f_2(1810)$, $f_2(1960)$, $f_2(2000)$,
$f_2(2240)$ and $f_2(2410)$ -- and with eleven states when adding one more
resonance $f_2(2020)$ which is needed in the combined analysis of data on
processes $p\overline{p}\to\pi\pi,\eta\eta,\eta\eta^\prime$ \cite{Anis05}.
The description with eleven states is practically the same as that with
ten resonances: the total
$\chi^2/\mbox{NDF}=156.617/(168-69)\approx 1.58$.

The parameters of the Breit--Wigner generators of the poles are shown
in Table~~\ref{tab:10f2-param} for the ten-states scenario and in
Table~~\ref{tab:11f2-param} for the eleven-states one.
\begin{table}[htb!]
\caption{The parameters of the Breit--Wigner forms for 10 $f_2$-states
(in MeV).} \vskip0.2truecm {
\def\arraystretch{1.0}
\begin{tabular}{|c|c|c|c|c|c|} \hline
{State} & ~$M_r$~ & $f_{r1}$ & $f_{r2}$ & $f_{r3}$ & $f_{r4}$ \\
\hline
{$f_2(1270)$} & 1275.3$\pm$1.8 & 470.8$\pm$5.4 & 22.4$\pm$4.6
& 201.5$\pm$11.4 & 90.4$\pm$4.76 \\
{$f_2(1430)$} & 1450.8$\pm$18.7 & 128.3$\pm$45.9 & 8.2$\pm$65
& 562.3$\pm$142 & 32.7$\pm$18.4 \\
{$f_2^{\prime}(1525)$} & 1535$\pm$8.6 & 28.6$\pm$8.3 & 41.6$\pm$160
& 253.8$\pm$78 & 92.6$\pm$11.5 \\
{$f_2(1600)$} & 1601.4$\pm$27.5 & 75.5$\pm$19.4 & 127$\pm$199
& 315$\pm$48.6 & 388.9$\pm$27.7 \\
{$f_2(1730)$} & 1723.4$\pm$5.7 & 78.8$\pm$43 & 107.6$\pm$76.7
& 289.5$\pm$62.4 & 460.3$\pm$54.6 \\
{$f_2(1810)$} & 1761.8$\pm$15.3 & 129.5$\pm$14.4 & 90.3$\pm$90
& 259$\pm$30.7 & 469.7$\pm$22.5 \\
{$f_2(1960)$} & 1962.8$\pm$29.3 & 132.6$\pm$22.4 & 65.4$\pm$94
& 333$\pm$61.3 & 319$\pm$42.6 \\
{$f_2(2000)$} & 2017$\pm$21.6 & 143.5$\pm$23.3 & 450.4$\pm$221
& 614$\pm$92.6 & 58.8$\pm$24 \\
{$f_2(2240)$} & 2207$\pm$44.8 & 136.4$\pm$32.2 & 166.8$\pm$104
& 551$\pm$149 & 375$\pm$114 \\
{$f_2(2410)$} & 2429$\pm$31.6 & 177$\pm$47.2 & 460.8$\pm$209
& 411$\pm$196.9 & 4.5$\pm$70.8 \\
\hline  \end{tabular}}
\label{tab:10f2-param}
\end{table}

The background parameters for the ten-states scenario are:
$\alpha_{11}=-0.07805$, $\alpha_{12}=0.03445$,
$\alpha_{10}=-0.2295$, $\beta_1=-0.0715$, $\gamma_1=-0.04165$,
$\beta_2=-0.981$, $\gamma_2=0.736$, $\beta_3=-0.5309$,
$\gamma_3=0.8223$.

\begin{table}[htb!]
\caption{The parameters of the Breit--Wigner forms for 11 $f_2$-states.}
\vskip0.3truecm {
\def\arraystretch{1.0}
\begin{tabular}{|c|c|c|c|c|c|} \hline
{State} & ~$M_r$~ & $f_{r1}$ & $f_{r2}$ & $f_{r3}$ & $f_{r4}$ \\
\hline
{$f_2(1270)$} & 1276.3$\pm$1.8 & 468.9$\pm$5.5 & 7.2$\pm$4.6
& 201.6$\pm$11.6 & 89.9$\pm$4.79 \\
{$f_2(1430)$} & 1450.5$\pm$18.8 & 128.3$\pm$45.9 & 8.2$\pm$63
& 562.3$\pm$144 & 32.7$\pm$18.6 \\
{$f_2^{\prime}(1525)$} & 1534.7$\pm$8.6 & 28.5$\pm$8.5 &
51.6$\pm$155 & 253.9$\pm$79 & 89.5$\pm$12.5 \\
{$f_2(1600)$} & 1601.5$\pm$27.9 & 75.5$\pm$19.6 & 127$\pm$190
& 315$\pm$50.6 & 388.9$\pm$28.6 \\
{$f_2(1730)$} & 1719.8$\pm$6.2 & 78.8$\pm$43 & 108.6$\pm$76.
& 289.5$\pm$62.6 & 460.3$\pm$54.5 \\
{$f_2(1810)$} & 1760$\pm$17.6 & 129.5$\pm$14.8 & 90.3$\pm$89.5
& 259$\pm$32 & 469.7$\pm$25.2 \\
{$f_2(1960)$} & 1962.2$\pm$29.8 & 132.6$\pm$23.3 & 62.4$\pm$91.3
& 331$\pm$61.5 & 319$\pm$42.8 \\
{$f_2(2000)$} & 2006$\pm$22.7 & 155.7$\pm$24.4 & 574.8$\pm$211
& 169.5$\pm$95.3 & 60.4$\pm$26.7 \\
{$f_2(2020)$} & 2027$\pm$25.6 & 50.4$\pm$24.8 & 128$\pm$190
& 441$\pm$196.7 & 58$\pm$50.8 \\
{$f_2(2240)$} & 2202$\pm$45.4 & 133.4$\pm$32.6 & 168.8$\pm$103
& 545$\pm$150.4 & 381$\pm$116 \\
{$f_2(2410)$} & 2387$\pm$33.3 & 175$\pm$48.3 & 462.8$\pm$211
& 395$\pm$197.7 & 24.5$\pm$68.5 \\
\hline
\end{tabular}}
\label{tab:11f2-param}
\end{table}
The background parameters for the eleven-states case are:
$\alpha_{11}=-0.0755$,
$\alpha_{12}=0.0225$, $\alpha_{10}=-0.2344$, $\beta_1=-0.0782$,
$\gamma_1=-0.05215$, $\beta_2=-0.985$, $\gamma_2=0.7494$,
$\beta_3=-0.5162$, $\gamma_3=0.786$.

\begin{figure}[htb]
\begin{center}
\hspace*{-0.2cm}
\includegraphics[width=0.8\textwidth,angle=0]{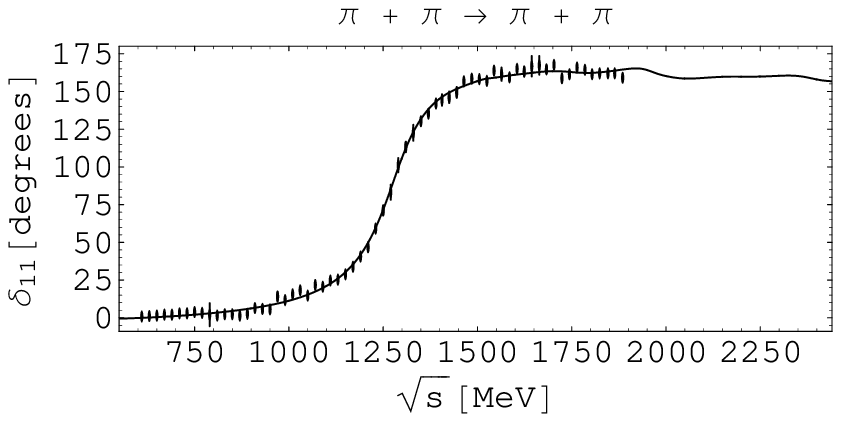}
\hspace*{-0.2cm}
\includegraphics[width=0.8\textwidth,angle=0]{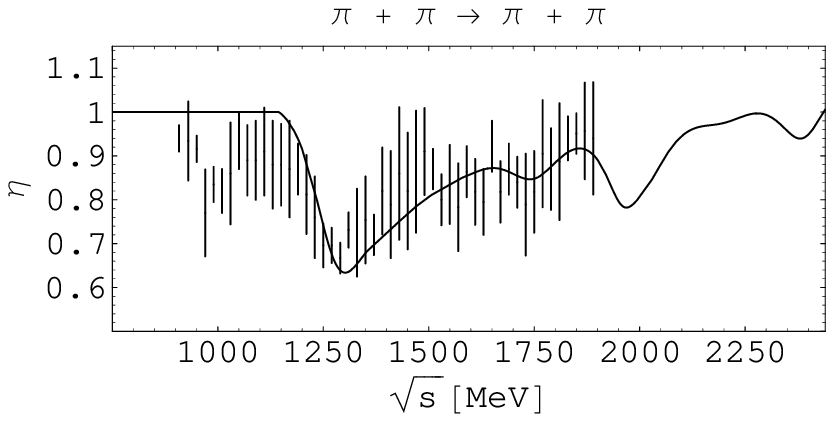}
\end{center}
\vspace*{-0.8cm} \caption{The phase shift and module of the
$D$-wave $\pi\pi$-scattering $S$-matrix element.}
\label{tens:ph-mod_pipi}
\end{figure}

\begin{figure}[htb]
\begin{center}
\hspace*{-0.2cm}
\includegraphics[width=0.8\textwidth,angle=0]{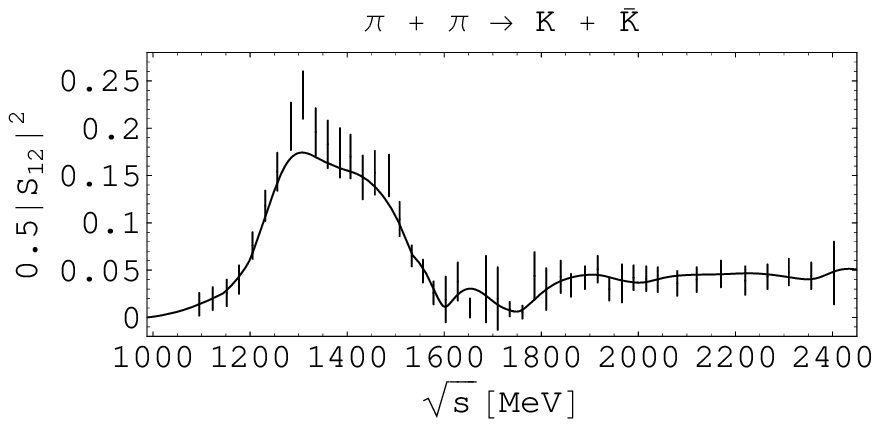}
\hspace*{-0.2cm}
\includegraphics[width=0.8\textwidth,angle=0]{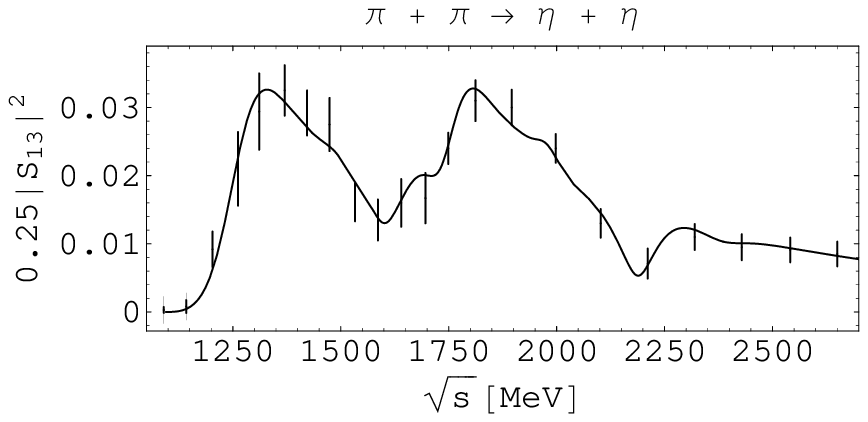}
\end{center}
\vspace*{-0.8cm} \caption{The squared modules of the $\pi\pi\to
K\overline{K}$ and $\pi\pi\to\eta\eta$ $D$-wave $S$-matrix elements.}
\label{tens:mod_K_eta}
\end{figure}

In the following we consider the eleven-states scenario (see
discussion in Section V). In Figures \ref{tens:ph-mod_pipi} and
\ref{tens:mod_K_eta} we demonstrate obtained energy dependences of
the analyzed quantities, compared with the experimental data.

In Table~\ref{tab:f2-poles} we show the poles in the complex energy plane
$\sqrt{s}$ (obtained using eqs.(\ref{d_{res}}) and (\ref{R_{ri}})), which
must be used for the calculation of the masses and widths of resonances.

\begin{table}[htb!]
\caption{The $f_2$-resonance poles on sheets II, IV, VIII, and XVI for
eleven states.
$\sqrt{s_r}={\rm E}_r-i\Gamma_r/2$~ in MeV is given.} {
\begin{tabular}{|c|cc|cc|cc|cc|}
\hline {} & \multicolumn{2}{c|}{II} & \multicolumn{2}{c|}{IV} &
\multicolumn{2}{c|}{VIII} & \multicolumn{2}{c|}{XVI} \\ \hline
{State} & ${\rm E}_r$ & $\Gamma_r/2$ & ${\rm E}_r$ & $\Gamma_r/2$
& ${\rm E}_r$ & $\Gamma_r/2$ & ${\rm E}_r$ & $\Gamma_r/2$ \\
\hline {$f_2(1270)$} & $1282\!\pm\!2.6\!$ & $67.5\!\pm\!4.2\!$ &
$1257\!\pm\!3.5\!$ & $99.6\!\pm\!3\!$ & $1277\!\pm\!3\!$ &
$73.4\!\pm\!4\!$ & $1264\!\pm\!3.4\!$ &
$98\!\pm\!3.5\!$ \\
{$f_2(1430)$} & $1425\!\pm\!48\!$ & $98.8\!\pm\!54\!$ &
$1421\!\pm\!49\!$ & $109\!\pm\!53\!$ & $1426\!\pm\!48\!$ &
$98\!\pm\!55\!$ & $1422\!\pm\!49\!$ &
$109\!\pm\!52\!$ \\
{$f_2^{\prime}(1525)$} & $1534\!\pm\!13\!$ & $24\!\pm\!28\!$ &
$1534\!\pm\!13\!$ & $23\!\pm\!9\!$ & $1534\!\pm\!13\!$ &
$17\!\pm\!29\!$
& $1534\!\pm\!13\!$ & $19.5\!\pm\!28\!$ \\
{$\!f_2(1600)\!$} & $1590\!\pm\!44\!$ & $80.5\!\pm\!34\!$ &
$1592\!\pm\!41\!$ & $74\!\pm\!34\!$ & $1600\!\pm\!41\!$ &
$23\!\pm\!35\!$ & $1601\!\pm\!40\!$ &
$9.4\!\pm\!35\!$ \\
{$f_2(1710)$} & $1710\!\pm\!12\!$ & $87\!\pm\!27\!$ &
$1711\!\pm\!11\!$ & $84\!\pm\!27\!$ & $1717\!\pm\!9.6\!$ &
$42.4\!\pm\!27\!$ & $1718\!\pm\!9\!$ &
$32\!\pm\!27\!$ \\
{$f_2(1810)$} & $1752\!\pm\!26\!$ & $79\!\pm\!15\!$ &
$1752\!\pm\!26\!$ & $84\!\pm\!15\!$ & $1757\!\pm\!25\!$ &
$50.6\!\pm\!15\!$ & $1758\!\pm\!25\!$ &
$36.5\!\pm\!15\!$ \\
{$f_2(1960)$} & $1958\!\pm\!43\!$ & $50\!\pm\!19\!$ &
$1957\!\pm\!43\!$ & $57\!\pm\!19\!$ & $1962\!\pm\!42\!$ &
$3.5\!\pm\!19\!$ & $1962\!\pm\!42\!$ &
$7.4\!\pm\!19\!$ \\
{$f_2(2000)$} & $2003\!\pm\!36\!$ & $84\!\pm\!62\!$ &
$2004\!\pm\!35\!$ & $68\!\pm\!64\!$ & $2003\!\pm\!35\!$ &
$82\!\pm\!64\!$ & $2002\!\pm\!36\!$ &
$95\!\pm\!62\!$ \\
{$f_2(2020)$} & $2025\!\pm\!39\!$ & $52\!\pm\!51\!$ &
$2026\!\pm\!38\!$ & $45.4\!\pm\!57\!$ & $2026\!\pm\!38\!$ &
$42.5\!\pm\!57\!$ & $2025\!\pm\!39\!$ &
$52\!\pm\!51\!$ \\
{$f_2(2240)$} & $2196\!\pm\!62\!$ & $103\!\pm\!54.5\!$ &
$2197\!\pm\!62\!$ & $98\!\pm\!55\!$ & $2202\!\pm\!61\!$ &
$24\!\pm\!57\!$ & $2201\!\pm\!62\!$ &
$45\!\pm\!57\!$ \\
{$f_2(2410)$} & $2385\!\pm\!49\!$ & $71\!\pm\!58\!$ &
$2387\!\pm\!47\!$ & $5.6\!\pm\!61\!$ & $2387\!\pm\!48\!$ &
$18.7\!\pm\!60\!$ & $2385\!\pm\!49\!$ &
$84\!\pm\!59\!$\\
\hline
\end{tabular}}
\label{tab:f2-poles}
\end{table}

Errors of the pole positions shown in Table~\ref{tab:f2-poles} are
estimated using a Monte Carlo method. In this method, the parameters $M_r$
and $f_{rj}$ are randomly generated using a normal distribution (Gaussian)
with the width given by the parameter error in Table~\ref{tab:11f2-param}.
Having generated the parameters, distributions (histograms for deviations
of the pole positions) for the real and imaginary parts of the pole
positions are evaluated and the standard deviations, which characterize
``widths'' of the distributions for the pole position, are calculated.

The masses $m_{res}$ and total widths $\Gamma_{tot}$ of states are
calculated from the pole positions using the denominator of the resonance
part of amplitude in the form
$$
T^{res}=\sqrt{s}\Gamma_{el}/(m_{res}^2-s-i\sqrt{s}\Gamma_{tot}).
$$
Then
\begin{equation}
m_{res}=\sqrt{{\rm E}_r^2+(\Gamma_r/2)^2},~~~
\Gamma_{tot}=\Gamma_r.
\end{equation}
The obtained values of the $m_{res}$ and $\Gamma_{tot}$ are shown
in Table \ref{tab:f2-mass}. It is clear that the values of these
quantities, calculated from the pole positions on various sheets,
mutually slightly differ; for the $f_2(2240)$ and
$f_2(2410)$, lying in the energy region where data are very
scanty, even considerably. We show only the values which
match best the corresponding values $M_r$ and the quantities
$\sum_{i=1}^Nf_{ri}^2/M_r$. The sheets on which the poles, used in
calculation of $m_{res}$ and $\Gamma_{tot}$, lie are also
indicated. If two sheets are indicated, the pole
positions on these sheets do not differ more than 1-1.5 MeV.
\begin{table}[htb!]
\caption{The masses and total widths of the $f_2$-resonances (all
in MeV)}. {
\begin{tabular}{|c|c|c|c|c|c|c|}
\hline {} & $f_2(1270)$ & $f_2(1430)$ & $f_2^{\prime}(1525)$ &
$f_2(1600)$ & $f_2(1710)$ & $f_2(1810)$ \\
\hline $\!m_{res}\!$ & $1268.0\!\pm\!3.4$ & $1425.5\!\pm\!49.2$ &
$1533.8\!\pm\!13.4$ & $1592.3\!\pm\!44.3$ & $1712.2\!\pm\!11.6$ &
$1753.8\!\pm\!25.6$ \\
$\!\Gamma_{tot}\!$ & $196.0\!\pm\!7.0$ & $218.6\!\pm\!105.4$ &
$48.4\!\pm\!56.0$ &
$161.0\!\pm\!68.6$ & $174.0\!\pm\!53.8$ & $167.6\!\pm\!29.4$ \\
Sheet  & XVI & IV, XVI & II, IV & II & II & IV \\
\hline\hline
{} & $f_2(1960)$ & $f_2(2000)$ & $f_2(2020)$ & $f_2(2240)$ & $f_2(2410)$ &  \\
\hline $\!m_{res}\!$ & $1958.0\!\pm\!42.9$ & $2004.0\!\pm\!36.3$ &
$2026.0\!\pm\!39.0$ & $2198.8\!\pm\!62.3$ & $2386.0\!\pm\!48.7$ &  \\
$\!\Gamma_{tot}\!$ & $113.6\!\pm\!37.0$ & $189.2\!\pm\!123.2$ &
$104.4\!\pm\!102.2$ & $205.6\!\pm\!109.0$ & $167.6\!\pm\!117.0$ &  \\
Sheet & IV & XVI & II, XVI & II & XVI &  \\ \hline
\end{tabular}}
\label{tab:f2-mass}
\end{table}

\section{Analysis of the isovector $F$-wave of $\pi\pi$
scattering}

In analysis of the $\pi\pi$-scattering data in the
$I^GJ^{PC}=1^+3^{--}$ sector by B.~Hyams et al. \cite{v-pipi},
we took into account that the dominant modes of decay of
the $\rho_3(1690)$ are $\pi\pi$, $4\pi$, $\omega\pi$,
$K\overline{K}$ and $K\overline{K}\pi$, and therefore have used
the four-channel Breit--Wigner forms in constructing the Jost matrix
determinant
$d(\sqrt{s-s_1},\sqrt{s-s_2},\sqrt{s-s_3},\sqrt{s-s_4})$ where
$s_1$, ..., $s_4$ are, respectively, the thresholds of the first four
channels given above. The resonance poles and zeros in the
$S$-matrix are generated by the Le~Couteur--Newton relation
\begin{equation}
S_{11}=d(-\sqrt{s-s_1},\cdots,\sqrt{s-s_4})/d(\sqrt{s-s_1},
\cdots,\sqrt{s-s_4})\;.
\end{equation}
The resonance part (\ref{d_{res}}) of the $d$-function has the form
\begin{equation}
d_{res}(s)=\prod_{r}
\left[M_r^2-s-i\sum_{j=1}^4\left(\sqrt{\frac{s-s_j}
{M_r^2-s_j}}\right)^7~R_{rj}~f_{rj}^2\right].
\end{equation}
The Blatt--Weisskopf factor for a particle with $J=3$ is
\begin{equation}
\!R_{rj}\!=\!\frac{15\!+\!3(\sqrt{\!M_r^2-s_j}~r_{rj})^2\!+
\!\frac{2}{5}(\sqrt{\!M_r^2-s_j}~r_{rj})^4\!+\!\frac{1}{15}
(\!\sqrt{M_r^2-s_j}~r_{rj})^6}
{15+3(\sqrt{s-s_j}~r_{rj})^2+\frac{2}{5}(\sqrt{s-s_j}~r_{rj})^4+
\frac{1}{15}(\sqrt{s-s_j}~r_{rj})^6}\!
\end{equation}
with radii of 0.927 fm in all channels as the result of the analysis.

The background part (\ref{d_B}) turned out to be elastic:
\begin{equation}
d_B=\exp\left[-i\left(\sqrt{\frac{s-4m_{\pi}^2}{s}}\right)^7 ~a_1
\right]\;
\end{equation}
where $a_1=-0.0138\pm0.0011$.

\begin{figure}[htb!]
\begin{center}
\hspace*{-0.2cm}
\includegraphics[width=0.7\textwidth,angle=0]{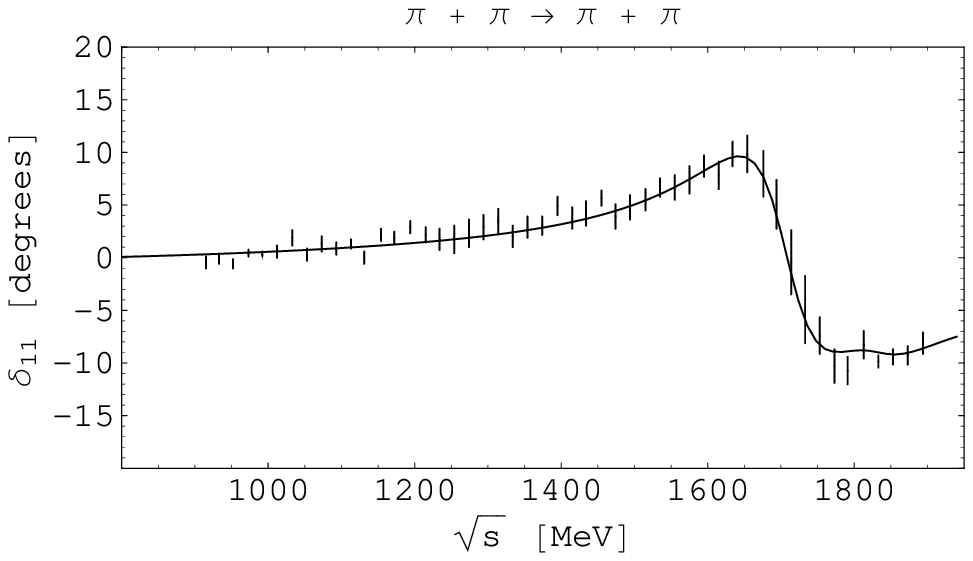}
\hspace*{-0.2cm}
\includegraphics[width=0.7\textwidth,angle=0]{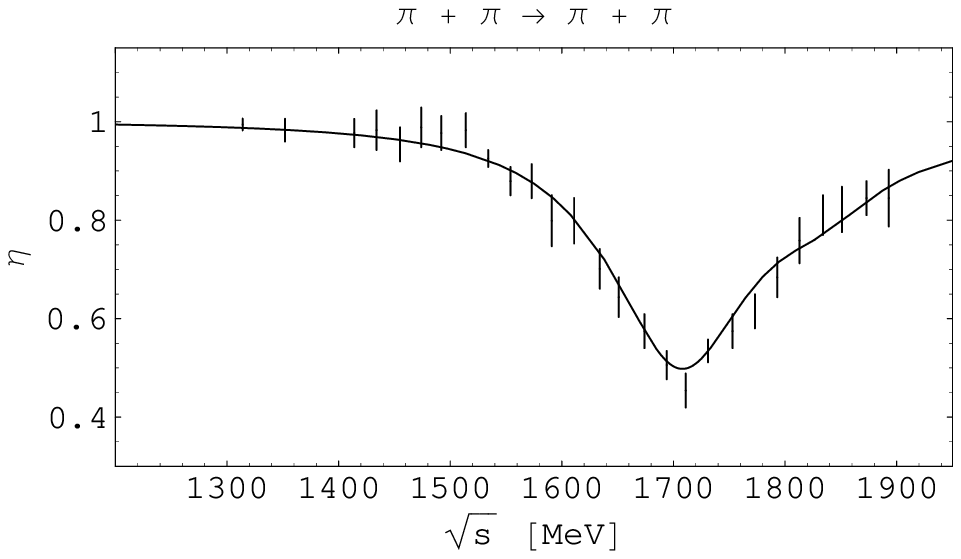}
\end{center}
\vspace*{-0.8cm} \caption{The phase shift and module of the
$\pi\pi$-scattering $F$-wave $S$-matrix element.}
\label{F:ph-mod_pipi}
\end{figure}

In the analysis we considered the cases with one and two resonances.
We obtained a good description in both cases: the total
$\chi^2/\mbox{NDF}\approx1$. In Figure \ref{F:ph-mod_pipi} we
show results of our fitting to the data for the case of two
resonances.

The obtained parameters of the Breit--Wigner forms and the
generated poles on the relevant sheets are shown in
Tables~\ref{tab:rho_BW-param} and \ref{tab:rho_3-poles},
respectively.
\begin{table}[htb!]
\caption{The parameters of the Breit--Wigner forms for two
$\rho_3$-like states (all in MeV).}
\def\arraystretch{1.0}
{\small
\begin{tabular}{|c|c|c|c|c|c|} \hline
{State} & ~$M_r$~ & $f_{r1}$ & $f_{r2}$ & $f_{r3}$ & $f_{r4}$ \\
\hline {$~\rho_3(1690)~$} & 1707.8$\pm$13.7 & 284.4$\pm$15.9 &
435.3$\pm$21.0 & 208.6$\pm$18.4 & 113.5$\pm$25 \\
{$~\rho_3(1950)~$} & 1833.5$\pm$28.6 & 96.3$\pm$18.3 &
331.8$\pm$28.0 & 297.7$\pm$16.5 & 110.4$\pm$28.3 \\
\hline  \end{tabular}} \label{tab:rho_BW-param}
\end{table}
\begin{table}[htb!]
\caption{The poles, generated by the Breit--Wigner forms on sheets
II, IV, VIII, and XVI. $\!\sqrt{s_r}={\rm E}_r-i\Gamma_r/2$~ in
MeV is given.}
\def\arraystretch{1.0}{\small
\begin{tabular}{|c|cc|cc|cc|cc|}\hline
{} {} & \multicolumn{2}{c|}{II} & \multicolumn{2}{c|}{IV} &
\multicolumn{2}{c|}{VIII} & \multicolumn{2}{c|}{XVI} \\ \hline
{State} & ${\rm E}_r$ & $\Gamma_r/2$ & ${\rm E}_r$ & $\Gamma_r/2$
& ${\rm E}_r$ & $\Gamma_r/2$ & ${\rm E}_r$ & $\Gamma_r/2$ \\
\hline {$~\rho_3(1690)~$} & $1705\!\pm\!5.6$ & $\!48\!\pm\!8\!$ &
$1707.6\!\pm\!4.5$ & $\!15.3\!\pm\!13\!$ & $1703.6\!\pm\!3.9$ &
$\!70\!\pm\!14\!$ & $1700.5\!\pm\!4.4$ &
$87.7\!\pm\!13.5$ \\
{$~\rho_3(1950)~$} & $1830.4\!\pm\!28$ & $\!55\!\pm\!14\!$ &
$1833.5\!\pm\!29$ & $\!0.0\!\pm\!22.7\!$ & $1833.5\!\pm\!27.5$ &
$\!11.7\!\pm\!15\!$ & $\!1831\!\pm\!24.3\!$ &
$\!53.3\!\pm\!22.3\!$ \\
\hline
\end{tabular}}
\label{tab:rho_3-poles}
\end{table}

Finally, in Table~\ref{tab:rho_3-mass} we show the mass and total
width of the $\rho_3(1690)$ and its branching ratios compared with
the average values from the PDG tables.
\begin{table}[htb!]
\caption{The parameters of the $\rho_3(1690)$ and its branching
ratios compared with the average values from the PDG tables.}
\vskip0.3truecm
\def\arraystretch{1.0}
{\small
\begin{tabular}{|c|c|c|c|c|c|c|c|} \hline
Scenario & $m_{res}\!$ [MeV] & $\Gamma_{tot}\!$ [MeV] &
$\!\Gamma_{\pi\pi}/\Gamma_{tot}\!$ &
$\!\Gamma_{\pi\pi}/\Gamma_{4\pi}\!$ &
$\!\Gamma_{K\overline{K}}/\Gamma_{\pi\pi}\!$ &
$\!\Gamma_{\omega\pi}/\Gamma_{4\pi}\!$ &
$\!\Gamma_{K\overline{K}}/\Gamma_{tot}\!$\\
\hline 1 state & $1703\!\pm\!4\!$ & $\!179\!\pm\!12\!$ &
$0.29\!\pm\!0.022\!$ & $0.472\!\pm\!0.097\!$ &
$0.146\!\pm\!0.06\!$ & $0.235\!\pm\!0.04\!$ &
$0.042\!\pm\!0.03\!$\\
\hline 2 states & $1702.7\!\pm\!4\!$ & $\!175\!\pm\!11\!$ &
$\!0.271\!\pm\!0.021\!$ & $\!0.427\!\pm\!0.096\!$ &
$\!0.159\!\pm\!0.045\!$ & $\!0.23\!\pm\!0.04\!$
& $\!0.043\!\pm\!0.032\!$ \\
\hline PDG & $1688.8\!\pm\!2.1\!$ & $\!160\!\pm\!10\!$ &
$0.243\!\pm\!0.013\!$ & $0.332\!\pm\!0.026\!$ &
$0.118^{+0.039}_{-0.032}\!$ & $0.23\!\pm\!0.05\!$ &
$0.013\!\pm\!0.0024\!$ \\
\hline  \end{tabular}} \label{tab:rho_3-mass}
\end{table}

\section{Discussion and conclusions }
\begin{itemize}
\item
In the $I^GJ^{PC}=0^+2^{++}$ sector, we carried out two analyses --
without and with the $f_2(2020)$. We did not obtain
$f_2(1640)$, $f_2(1910)$, $f_2(2150)$ and $f_2(2010)$, however, we
saw $f_2(1430)$ and $f_2(1710)$ which are related to the
statistically-valued experimental points.
\item
Usually one assigns the states $f_2(1270)$ and $f_2^{\prime}(1525)$
to the first tensor nonet. One could assign the $f_2(1600)$ and
$f_2(1710)$ states to the second nonet, though the isodoublet
member is not discovered yet. If $a_2(1730)$ is the isovector of
this octet and if $f_2(1600)$ is almost its eighth component,
then from the GM-O formula
$$M_{K_2^*}^2=\frac{1}{4}(3M_{f_2(1600)}^2+M_{a_2(1730)}^2),$$
one would expect this isodoublet mass at about 1633~MeV. In the
relation for masses of nonet
$$M_{f_2(1600)}+M_{f_2(1710)}=2M_{K_2^*(1633)},$$
the left-hand side is only by 1.2\% larger than the right-hand one.

In Ref.~\cite{Karn-00}, one has observed the strange isodoublet
in the mode $K_s^0\pi^+\pi^-$ with yet indefinite remaining quantum
numbers and the mass $1629\pm7$ MeV. This state could be the
tensor isodoublet of the second nonet.
\item
The states $f_2(1963)$ and $f_2(2207)$ together with the
isodoublet $K_2^*(1980)$ could be put into the third nonet. Then
in the relation for masses of nonet
$$M_{f_2(1963)}+M_{f_2(2207)}=2M_{K_2^*(1980)},$$
the left-hand side is only by 5.3\% larger than the right-hand one.

If one considers $f_2(1963)$ as the eighth component of octet, the
GM-O formula gives $M_{a_2}=2030$ MeV. This value coincides with
that for the $a_2$ meson obtained in analysis \cite{Anis2-01}. This
state is interpreted \cite{Anis05} as the second radial excitation
of the $1^-2^{++}$ state based on consideration of the
$a_2$ trajectory on the $(n,M^2)$ plane where $n$ is the radial
quantum number of the $q{\bar q}$ state.
\item
As to the $f_2(2000)$, the presence of the $f_2(2020)$ in the
analysis with eleven resonances helps to interpret $f_2(2000)$ as
the glueball. In the case of ten resonances, the ratio of the
$\pi\pi$ and $\eta\eta$ widths is in the limits obtained in
Ref. \cite{Anis05} for the tensor glueball on the basis of the
$1/N_c$-expansion rules. However, the $K\overline{K}$ width is too
large for the glueball.
In both, practically the same, descriptions of the processes,
the parameters of $f_2$ states do not differ too much, except for
the $f_2(2000)$ and $f_2(2410)$. The mass of the latter has
decreased by about 40 MeV but the $K\overline{K}$ width of the
former has changed significantly. Now all the obtained ratios
of the partial widths are in the limits corresponding to the
glueball.\\
The question of interpretation of the $f_2(2020)$ and $f_2(2410)$
is open.
\item
Finally we have $f_2(1430)$ and $f_2(1710)$ which are neither
$q{\bar q}$ states nor glueballs. Since one predicts that masses
of the lightest $q{\bar q}g$ hybrids are bigger than those of
lightest glueballs, these states might be the 4-quark ones. Then
for the isodoublet mass of the corresponding nonet, we would
expect the value 1570-1600 MeV. For now we do not know any
experimental indications for the tensor isodoublet of that mass.
However, in the known experimental spectrum of the $K_2^*$ family,
there is a 500-MeV unoccupied gap from 1470 to 1970~MeV
\cite{PDG10}, except for the above work \cite{Karn-00}. Moreover,
as one can see in the PDG tables on the $a_2(1700)$ listing, widths
of the observed isovector tensor states in the 1660-1775-MeV interval
differ by the factor 2-3, i.e., the states possess various
properties. For example, the broad state with mass $1702\pm7$~MeV
and width $417\pm19$~ MeV, observed in ${\bar p}p\to\eta\eta\pi^0$
\cite{Uman-06}, might be the isovector member of the corresponding
four-quark nonet.

Of course, an assumption of this possibility presupposes an
existence of the scalar tetraquarks at lower energies
\cite{4quark} which are not seen in our analysis
\cite{SBKN-PRD10}. One can argue that these states are a part of
the background due to their very large widths.
\item
The analysis of the $F$-wave $\pi\pi$ scattering data by B.~Hyams
et al. \cite{v-pipi} indicates that, except for the known
$\rho_3(1690)$ (in our analysis $m_{res}\!\approx\!1703$~MeV,
$\Gamma_{tot}\!\approx\!175$~MeV), there might be one more
state lying above 1830~MeV. Since the $\pi\pi$ scattering data
above 1890~MeV are absent, it is impossible to say something
conclusive on parameters of this state. However, the $\rho_3(1950)$
does not contradict to the data but instead improves
a little bit the obtained parameters of the $\rho_3(1690)$ and
its branching ratios when
comparing them with the PDG tables \cite{PDG10}.

\end{itemize}

\begin{acknowledgments}
The work has been supported in part by the RFBR grant
10-02-00368-a, the Votruba-Blokhintsev Program for Cooperation of
the Czech Republic with JINR (Dubna), the Grant Agency of the
Czech Republic (Grant No.202/08/0984), the Slovak Scientific Grant
Agency (Grant VEGA No.2/0034/09), the Bogoliubov-Infeld
Program for Cooperation of Poland with JINR (Dubna) and the Grant Program
of Plenipotentiary of Slovak Republic at JINR.
\end{acknowledgments}

\end{document}